\documentclass[twocolumn,aps,prl,superscriptaddress,preprintnumbers,bibnotes]{revtex4-1}
\usepackage{graphicx} 
\usepackage{amsfonts} 
\usepackage{braket,tikz}
\usepackage{xcolor}
\usepackage{soul}
\usepackage{amsmath}
\usepackage{mathtools}
\usepackage{siunitx}
\usepackage{xcolor}
\usepackage{hyperref}  
\hypersetup{
    colorlinks = true,
    linkcolor  = blue,
    citecolor  = blue,
    urlcolor   = blue,
    hidelinks  
}
\usepackage{cleveref}
\crefname{equation}{}{}        
\Crefname{equation}{}{}        

\newcommand{\comment}[1]{\iffalse {#1} \fi}

\newcommand{\CUAaddress}{Harvard-MIT Center for Ultracold Atoms, Cambridge, Massachusetts 02138, USA}
\newcommand{\HarvardPhysicsAddress}{Department of Physics, Harvard University, Cambridge, Massachusetts 02138, USA}
\newcommand{\HarvardChemistryaddress}{Department of Chemistry and Chemical Biology, Harvard University, Cambridge, Massachusetts 02138, USA}
\begin{document}
\title{Multi-Qubit Stabilizer Readout on a Dual-Species Rydberg Array}
\author{Yu~Wang}\thanks{These authors contributed equally to this work.}

 \affiliation{\HarvardChemistryaddress} 
 \affiliation{\HarvardPhysicsAddress}
 \affiliation{\CUAaddress}

 \author{Ryan~Cimmino}\thanks{These authors contributed equally to this work.}
 \affiliation{\HarvardChemistryaddress} 
 \affiliation{\HarvardPhysicsAddress}
 \affiliation{\CUAaddress}

 \author{Kenneth~Wang}
 \affiliation{\HarvardChemistryaddress} 
 \affiliation{\HarvardPhysicsAddress}
 \affiliation{\CUAaddress}

 \author{Santiago~Lopez}
 \affiliation{\HarvardChemistryaddress} 
 \affiliation{\HarvardPhysicsAddress}
 \affiliation{\CUAaddress}

 \author{Jeffrey~Li}
 \affiliation{\HarvardChemistryaddress} 
 \affiliation{\HarvardPhysicsAddress}
 \affiliation{\CUAaddress}

 \author{Jin~Ming~Koh}
 \affiliation{\HarvardPhysicsAddress}
 \affiliation{\CUAaddress}

 \author{Jonathan~N.~Hall\'en}
 \affiliation{\HarvardPhysicsAddress}
 \affiliation{\CUAaddress}
 \affiliation{Department of Physics, Boston University, Boston, Massachusetts 02215, USA}

 \author{Anne~Matthies}
 \affiliation{\HarvardPhysicsAddress}
 \affiliation{\CUAaddress}

 \author{Norman~Y.~Yao}
 \email{nyao@fas.harvard.edu}
 \affiliation{\HarvardPhysicsAddress}
 \affiliation{\HarvardChemistryaddress} 
 \affiliation{\CUAaddress}

 \author{Kang-Kuen~Ni}
 \email{ni@chemistry.harvard.edu}
 \affiliation{\HarvardChemistryaddress} 
 \affiliation{\HarvardPhysicsAddress}
 \affiliation{\CUAaddress}



\begin{abstract}
The ability to locally control and measure subsets of ancilla qubits in an efficient and crosstalk-free manner is a key ingredient in quantum error correction (QEC).
Dual-species neutral atom arrays offer an ideal implementation of these capabilities, enabling independent state preparation, manipulation, and detection on each species. 
In this work, we realize such a dual-species Rydberg array of Na and Cs atoms trapped in co-localized 2D optical tweezer arrays, using Na as an ancilla to measure stabilizers of surrounding Cs data qubits.
We identify the finite interspecies Rydberg-Rydberg interaction strength as a practical obstacle to high-fidelity multi-body entanglement and show that, by tuning the Rabi frequency and the detuning of the Rydberg driving field, the resulting geometric phase error can be compensated. This yields a protocol for simultaneous, non-destructive, \emph{in situ} stabilizer readout of multiple data qubits via global pulses alone. 
Using this protocol, we demonstrate non-destructive measurement of Pauli-Z stabilizers on four-qubit Cs plaquettes via a single global Rydberg pulse sequence.
Our results demonstrate dual-species tweezer arrays as a promising route towards scalable QEC and open the door to new quantum control protocols leveraging both interspecies and intraspecies interactions.

\end{abstract}
\maketitle

\section{Introduction}
Quantum error correction (QEC) is a prerequisite for scalable quantum computation~\cite{Shor1995QC, Steane1996, Gottesman:1997zz, preskill2025beyond, Campbell2017}.
It relies on two complementary capabilities: (i) the ability to repeatedly measure multi-qubit operators (i.e.~stabilizers) and (ii) the ability to perform those measurements without perturbing the logical quantum information~\cite{beverland2022assessingrequirementsscalepractical, Gidney2021howtofactorbit, Litinski2019gameofsurfacecodes, Xu2024, Fowler2012SurfaceCodes,google2025quantum,lacroix2025scaling}. Owing, in part, to their flexible reconfigurability, neutral atom arrays have recently emerged as a promising platform for quantum computing and are particularly well-suited to the former: coherent atom transport allows ancilla qubits to be brought into contact with different sets of data qubits for multi-qubit stabilizer measurements~\cite{Saffman2010, bluvstein2022quantum, Cong2022}. 
To ensure the latter, neutral atom  platforms typically leverage a zoned architecture, where readout occurs in a spatially distinct region~\cite{bluvstein2022quantum}.
In such architectures, the mid-circuit transport of atoms into the readout zone --- which is typically two orders of magnitude slower than two-qubit gates --- dominates the duration of an error correction cycle and fundamentally limits scalability~\cite{Bluvstein2024, cain2026shorsalgorithmpossible10000}. 
Progress on this front therefore calls for new  primitives that enable parallel, \emph{in situ} stabilizer readout. 
%

\begin{figure*}[!htbp]
   \centering
    \includegraphics[width=\linewidth]{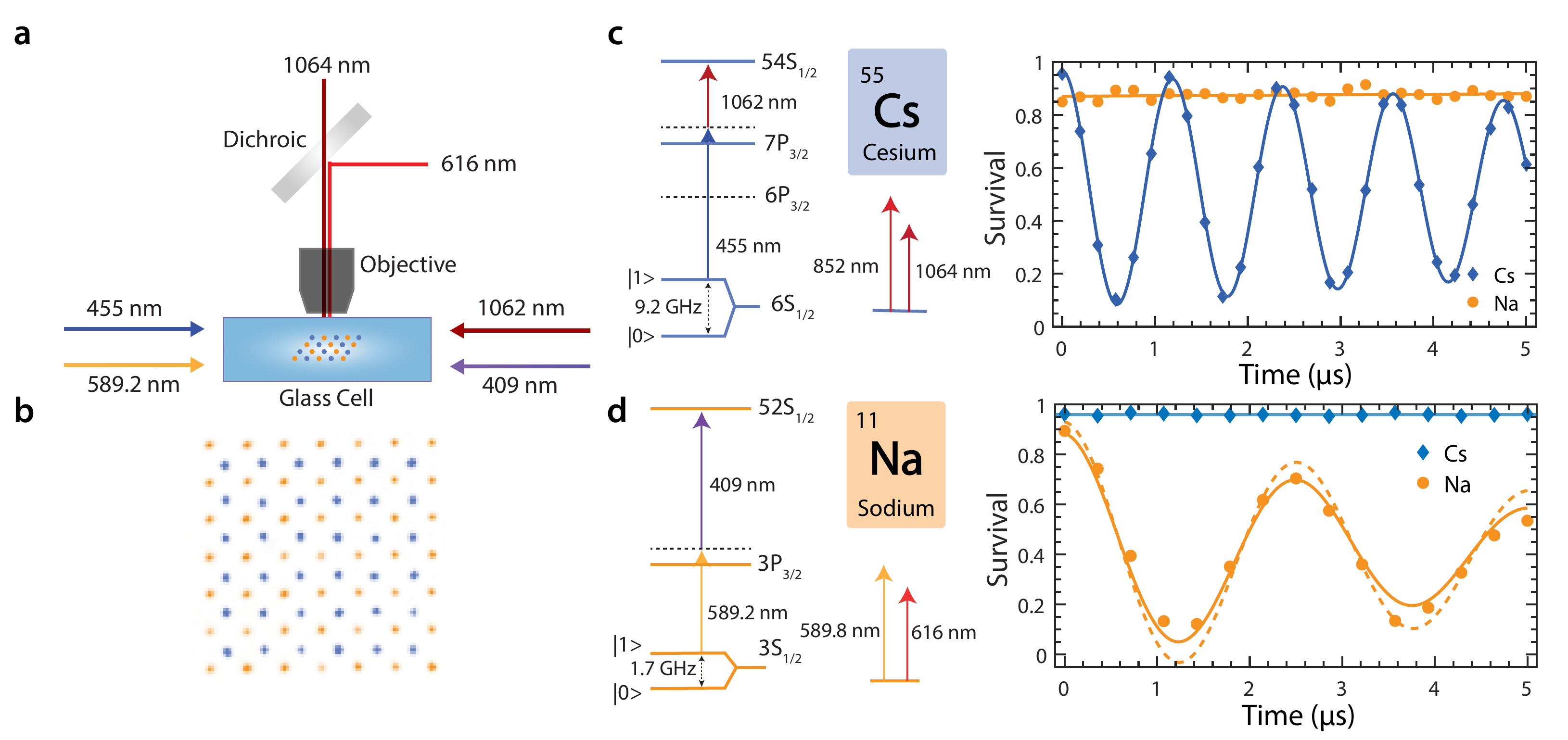}
    \caption{{\bf Experimental setup.} \textbf{(a)} The tweezer trapping beams are focused through the same objective to generate SLM tweezer arrays. Rydberg beams are combined (1062 nm with 409 nm, and 455 nm with 589 nm) for counter-propagating two-photon Rydberg excitation of each species. \textbf{(b)} A typical 2D array of Na (orange) and Cs (blue) atoms. \textbf{(c, left)} Select level structure of Cs showing states used in two-photon Rydberg excitation, as well as the hyperfine qubit ground states, the trapping wavelength, and the cooling/imaging wavelength. \textbf{(c, right)} Single-atom ground-Rydberg Rabi oscillations of Cs while nearby Na atoms are not perturbed. Error bars are smaller than the size of data points. \textbf{(d)} Analogous level-structure and ground-Rydberg Rabi oscillations of Na. The dashed line represents the oscillation after subtraction of the scattering background (see Supplementary Material). The Cs survival (blue) stays unperturbed. } 
    \label{fig:Exp_scheme}
\end{figure*}

In this work, we address this emerging bottleneck by introducing and demonstrating simultaneous, non-destructive stabilizer readout
on a dual-species ($\text{Na}$-$\text{Cs}$) Rydberg array. 
Our approach leverages two defining features of the dual-species platform \cite{Zeng2017,  Liu2018, doi:10.1126/science.ade5337, Anand2024,  Miles2026QubitSyndrome}. First, the spectroscopic distinction between the two species enables independent control---species-selective state preparation, manipulation, and detection---under global laser driving, eliminating the need for spatially separated operation zones. Second, the freedom to independently choose the Rydberg state of each species allows us to engineer strong interspecies interactions, enabling a single ancilla atom to 
interact with multiple surrounding data qubits through global Rydberg pulses alone, without the need for atom rearrangement or sequential pairwise gates \cite{jandura2026surfacecodestabilizermeasurements}. 

We demonstrate three main results. First, we establish the basic building blocks of the dual-species platform by realizing coherent ground--Rydberg excitation of each species and confirming species-selective driving via global controls.
Second, we identify the finite strength of interspecies Rydberg interactions as a generic obstacle to dual-species stabilizer readout, and implement a detuning-compensated protocol that relaxes the strict-blockade requirement. 
Finally, utilizing this protocol, we perform  a proof-of-principle demonstration of parallel, four-atom stabilizer readout.
In particular, we arrange data atoms (${}^{133}\text{Cs}$) on four-qubit plaquettes with an ancilla atom (${}^{23}\text{Na}$) at the center and implement the stabilizer readout using a single, global Rydberg pulse sequence. 

\section{Dual-species Rydberg array}

Our experiment employs Na and Cs atoms trapped in two co-localized optical tweezer arrays, each generated by a dedicated spatial light modulator (SLM) [Fig.~\ref{fig:Exp_scheme}\textbf{a, b}].
Throughout this work, we use Cs as the data qubit species and Na as the ancilla species. Quantum information in the Cs data register is encoded in the hyperfine stretched states $|0\rangle = |6S_{1/2},\ F = 3,\ m_F = 3\rangle$ and $|1\rangle = |6S_{1/2},\ F = 4,\ m_F =4\rangle$. Na atoms serve as ancillas for stabilizer readout and are initialized in $|g\rangle = |3S_{1/2},\ F = 2,\ m_F = 2\rangle$. Here, $F$ denotes the total angular momentum quantum number and $m_F$ the corresponding magnetic quantum number. To mediate interspecies interactions, we excite both species to Rydberg states $|r\rangle$ via two-photon transitions: for Cs, the $|1\rangle\to |r\rangle$ transition is driven through $|7P_{3/2}\rangle$~\cite{doi:10.1126/science.adq0278}, and for Na the $|g\rangle\to |r\rangle$ transition is driven through $|3P_{3/2}\rangle$ (Fig.~\ref{fig:Exp_scheme}\textbf{c,d}).
%

A key feature of the dual-species platform is that the Rydberg state of each species can be chosen independently, giving control over both the interspecies and intraspecies van der Waals interactions. In this work we choose $|r\rangle_\text{Cs} = |54S_{1/2},\ J=1/2,\ m_{J}=1/2\rangle$ and $|r\rangle_\text{Na} = |52S_{1/2},\ J=1/2,\ m_{J}=1/2\rangle$. We characterize species-selective control by driving coherent ground-Rydberg Rabi oscillations on each species independently. In these measurements, atoms excited to the Rydberg state are anti-trapped and appear as loss, while atoms remaining in the ground state survive and are imaged. Fig.~\ref{fig:Exp_scheme}\textbf{c} and \ref{fig:Exp_scheme}\textbf{d} show typical Rabi oscillations for Cs and Na, respectively. Crucially, when the global Rydberg beams for one species are applied, the other species remains unperturbed, directly demonstrating species-selective excitation under global driving.

\section{Two-atom stabilizer readout}
\label{section2}
\begin{figure*}[!htbp]
   \centering
    \includegraphics[width= 1\linewidth]{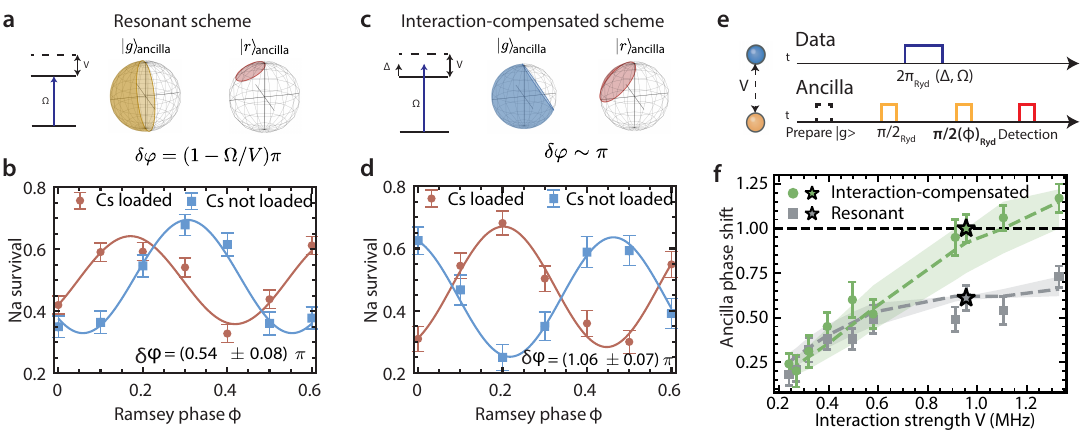}
        \caption{{\bf Interaction-compensated ancilla readout.} \textbf{(a,c)} The phase accumulated when the ancilla atom is in the ground state (yellow/blue, left) and Rydberg state (red, right).
    In \textbf{(a)}, the relative phase on the ancilla $\delta\varphi<\pi$ due to imperfect blockade. In \textbf{(c)}, at certain finite detunings $\Delta$ and Rabi frequencies $\Omega$, one can achieve $\delta\varphi\sim\pi$ (see main text).  Error bars in fitted $\delta\varphi$ include $1-\sigma$ bootstrap error and fitting error. \textbf{(b,d)} Ramsey fringe readout of $\delta\varphi$ for resonant scheme (a) and the interaction-compensated scheme (c), at $V = 1.1$ MHz as indicated by the black stars in (f). The compensated scheme exhibits a $\pi$ phase shift, while the resonant scheme shows a phase shift smaller than $\pi$ due to the finite interaction strength. \textbf{(e)} The experimental sequence implemented to measure $Z_\text{data}$, including state preparation, Rydberg laser pulses, and ancilla detection.  \textbf{(f)} Comparison of measured ancilla phase $\delta\varphi$ in units of $\pi$ for various interspecies interaction strengths using the resonant scheme (grey square) and the interaction-compensated scheme (green circle). The dashed lines show simulation results accounting for only the finite interaction, with the shaded region denoting $\pm 20\%$ interaction strength deviation. \textbf{(f, inset)} The pairwise array configuration used in this data. }
    \label{fig:twoatom}
\end{figure*}

\subsection{Stabilizer readout protocol on a dual-species array}
In the stabilizer formalism of QEC, logical states of an error-correcting code are defined as the simultaneous eigenstates of a set of stabilizer operators \cite{Gottesman:1997zz}, which are expressed as tensor products of single-qubit Pauli operators. 
For example, the paradigmatic surface code calls for the measurement of weight-four stabilizers of the form $Z_1Z_2Z_3Z_4$ on interleaved plaquettes of four physical qubits \cite{Litinski2019gameofsurfacecodes}. 
In practice, in order to ensure that the encoded logical state is unperturbed, this measurement often amounts to performing sequential two-qubit gates between an ancilla and each data qubit within the stabilizer, followed by measurement of the ancilla.
As discussed above, in a typical single-species neutral-atom architecture, these ancilla operations are distributed across spatially distinct preparation, entanglement, and readout zones \cite{bluvstein2022quantum}. 
In addition to the time overhead of rearrangement between these zones, suppression of crosstalk between data and ancilla qubits calls for a combination of shielding beams and local Raman addressing during ancilla preparation and measurement steps \cite{Bluvstein2026FaultTolerant}. 
Moreover, ancilla coherence must be preserved over tens to hundreds of milliseconds, requiring clock-state hyperfine encoding.

By contrast, the natural spectroscopic separation between species on our dual-species platform enables state preparation, gate implementation, and readout of the ancilla to be performed entirely \emph{in situ}, making this an ideal setting for stabilizer readout. 
Thus, atom transport can be eliminated, no longer requiring the ancilla to be encoded in the hyperfine clock states. 
Indeed, here, we implement a protocol in which the ancilla remains entirely within the ground-Rydberg manifold ($|g\rangle$, $|r\rangle$), and is simultaneously entangled with all data qubits through a single global operation.

%
%
%

%
%
Our protocol (Fig.~\ref{fig:twoatom}\textbf{e}) relies on a modified form of the $\pi-2\pi-\pi$ gate first described in \cite{Jaksch2000Gate}. The ancilla is initialized in $|g\rangle$ and pulsed to $(|g\rangle+i|r\rangle)/\sqrt{2}$ with a ground-Rydberg $\pi/2$ pulse.
Next, the data atom undergoes a ground-Rydberg $2\pi$ pulse on the $|1\rangle\to|r\rangle$ transition. In the limit of full interspecies blockade, a resonant $2\pi$ pulse imparts a phase $|1\rangle\to -|1\rangle$ only when the ancilla is in $|g\rangle$, so the ancilla acquires a relative phase conditioned on the state of the data qubit before readout. Finally, we perform a second $\pi/2$ pulse on the ancilla. Depending on the relative phase imparted by the data atom from its $2\pi$ pulse, this maps the ancilla to either $|g\rangle$ or $|r\rangle$:
\begin{align}
\label{eq1}
|0\rangle_\text{data}|g\rangle_\text{ancilla} &\to |0\rangle_\text{data}|r\rangle_\text{ancilla},\\
\label{eq2}
    |1\rangle_\text{data}|g\rangle_\text{ancilla} &\to |1\rangle_\text{data}|g\rangle_\text{ancilla}.
\end{align}
By detecting the ancilla qubit, the value of $ Z_\text{data}$ can be inferred directly from the survival of the ancilla without the need for additional rearrangement, hyperfine gate operations, or state-selective imaging. 
%
This protocol extends naturally to the setting of more than one data atom, as we explore in detail below. An additional $\pi$ phase is then acquired for each data atom in $|1\rangle$.


We first study this protocol for the case of two atoms. We prepare an array of Na and Cs atom pairs (Fig.~\ref{fig:twoatom}\textbf{f} inset) with Na ancilla atoms prepared in $|g\rangle$ and Cs data atoms prepared in $|1\rangle$, seeking to use the Na ancilla to non-destructively infer $Z_\text{data}$ on the single Cs data atom. 
This explicitly probes the behavior of the protocol for input states $|1\rangle_\text{data}$ (Eq.~\ref{eq2}); the case of $|0\rangle_\text{data}$ (Eq.~\ref{eq1}) is inferred from the behavior when a data atom is not loaded. We separately confirm (see Supplementary Material) that the gate performs identically when (a) data atoms are loaded in $|0\rangle$ and (b) data atoms are not loaded at all. 

The data atom $2\pi$ pulse is driven resonantly ($\Delta = 0$) at $\Omega_\text{data} = 918$ kHz.  For ancilla readout, we vary the phase $\phi$ of the final ancilla $\pi/2$ pulse relative to the initial $\pi/2$ pulse. The resulting Ramsey fringe, showing the return probability to $|g\rangle$ as a function of $\phi$, is plotted in Fig.~\ref{fig:twoatom}\textbf{b} for interspecies interaction strength $V = 1.1$ MHz. The case of a Cs data atom loaded in $|1\rangle$ is shown in red, and the case of no data atom loaded in blue. If Eqs.~\ref{eq1} and ~\ref{eq2} hold, the presence of a data atom in $|1\rangle$ should shift the ancilla fringe by $\delta\varphi = \pi$. We instead measure only $\delta\varphi = (0.54\pm0.08)\pi$ as shown in Fig.~\ref{fig:twoatom}\textbf{b}. 
%


\begin{figure*}[!ht]
   \centering
    \includegraphics[width =.9\linewidth]{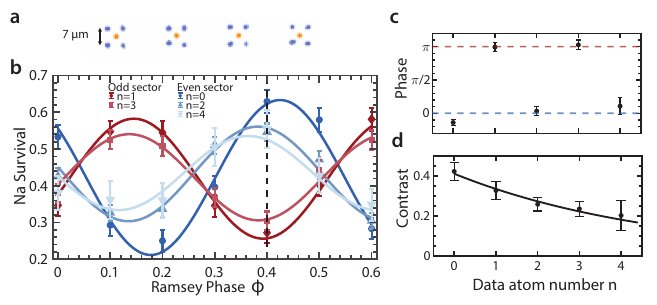}
    \caption{\textbf{Stabilizer measurement on a four-data-qubit plaquette.} \textbf{(a)} Fluorescence image of an array of plaquettes, each comprised of one central Na ancilla atom surrounded by four Cs data atoms. \textbf{(b)} Ramsey fringes measured on the Na ancilla for different numbers of loaded Cs atoms prepared in $|1\rangle$, $n=0$ to $4$, using the interaction-compensation protocol. The data separate into even-parity (blue) and odd-parity (red) sectors. Error bars denote statistical uncertainties. The dashed line marks the maximum discrimination between the two sectors at about $\phi = 0.4$.
    \textbf{(c)} Extracted Ramsey phase as a function of data-atom number $n$, showing an approximately $\pi$ phase difference between the even- and odd-parity sectors. Dashed lines are guides to the eye. \textbf{(d)} Fitted Ramsey contrast versus data-atom number $n$. The line shows an exponential decay with a readout fidelity of 0.82(3) per data qubit. Error bars include both $1-\sigma$ bootstrap and fitting uncertainties.}
    \label{fig_plaq}
\end{figure*}

\subsection{Stabilizer readout beyond full blockade}

This reduction of the phase shift can be understood as a consequence of being well outside of the Rydberg blockade regime defined by $\Omega_\text{data}\ll V$.  
When $V\sim\Omega_\text{data}$, the data atom undergoes population and phase dynamics even when the ancilla is in $|r\rangle$ as shown in Fig.~\ref{fig:twoatom}\textbf{a}. 
The relative phase acquired by the ancilla superposition is then less than $\pi$. 
We verify this by comparing gate performance at a range of interspecies interaction strengths $V$ with two-atom simulations accounting for $\Omega_\text{data}$ and $V$ (Fig.~\ref{fig:twoatom}\textbf{f}, grey dashed); the observed ancilla phase shifts are in good agreement with simulation. The value of $\delta\varphi$ increases towards $\pi$ as the interaction $V$ increases, corresponding to stronger interspecies blockade.

%
%

Rather than seeking to completely suppress the residual dynamics with further enhanced interaction, we instead demonstrate a modified protocol that achieves $\delta\varphi \approx \pi$ without requiring perfect interspecies blockade \cite{Shi2017, cole2026asymmetricfastrydberggate}.
%
In this interaction-compensated protocol, the data atom undergoes a closed trajectory on the Bloch sphere and acquires a geometric phase for both ancilla states $|g\rangle$ and $|r\rangle$, but the detuning $\Delta$ and Rabi frequency $\Omega_{\rm data}$ of the middle $2\pi$ pulse are chosen such that the difference of these phases is $\delta\varphi = \pi$ as shown in Fig.~\ref{fig:twoatom}\textbf{c} (see discussion in Supplementary Material). 
In particular, choosing $\Delta = V/2$ and $\Omega_{\rm data} = \sqrt{3}V/2$ completely compensates for the finite interspecies interaction strength. 
To confirm this, we attempt the interaction-compensated readout protocol with pulse parameters chosen for $V = 1.1$ MHz, and find that the desired $\delta\varphi = \pi$ is indeed achieved at finite $V$ (Fig.~\ref{fig:twoatom}\textbf{d}). 
We verify this picture across a range of interspecies interaction strengths, finding good agreement with simulation throughout (Fig.~\ref{fig:twoatom}\textbf{f}, green). This agreement validates the interaction-compensated protocol as a practical tool and establishes that stabilizer readout is achievable at the finite interaction strengths $V\sim 1$ MHz realizable on many-atom plaquettes, to which we now turn.

\section{Stabilizer readout on a four-data-qubit plaquette}


We now prepare an array of five-atom plaquettes, each consisting of a central Na ancilla surrounded by four Cs data qubits (Fig.~\ref{fig_plaq}\textbf{a}). We exploit the stochastic nature of the loading process to realize plaquettes with different numbers of data qubits $n = 0, 1, 2,3, 4$ in the $\ket{1}$ state, allowing us to probe all configurations in a single experiment. As above, we infer the behavior of the protocol for data atoms in $|0\rangle$ by using non-loaded atoms as a proxy. 
Applying the same stabilizer readout protocol to measure $Z_1 Z_2 Z_3 Z_4$ on the data qubits, we expect to observe two sets of Ramsey fringes: one set for the even sector $Z_1 Z_2 Z_3 Z_4 = +1$  ($n= 0, 2, 4$), and a second set with a relative $\pi$ phase shift for the odd sector $Z_1 Z_2 Z_3 Z_4 = -1$ ($n = 1,3$).

This expectation is indeed borne out by the data in Fig.~\ref{fig_plaq}\textbf{b}. The separation into even (blue) and odd (red) sectors enables a straightforward stabilizer readout by fixing the Ramsey phase at the operating point where the even- and odd-sector signals are maximally discriminated, indicated by the vertical dashed line at $\phi = 0.4$ in Fig.~\ref{fig_plaq}\textbf{b}. Ancilla survival (loss) indicates a $+1$ ($-1$) measurement outcome for the stabilizer $Z_1 Z_2 Z_3 Z_4$.

To fully validate this stabilizer readout beyond specially chosen input configurations, we perform the measurement for a more generic initial state (again fixing $\phi = 0.4$). 
\begin{figure}
    \centering
    \includegraphics[width=1\linewidth]{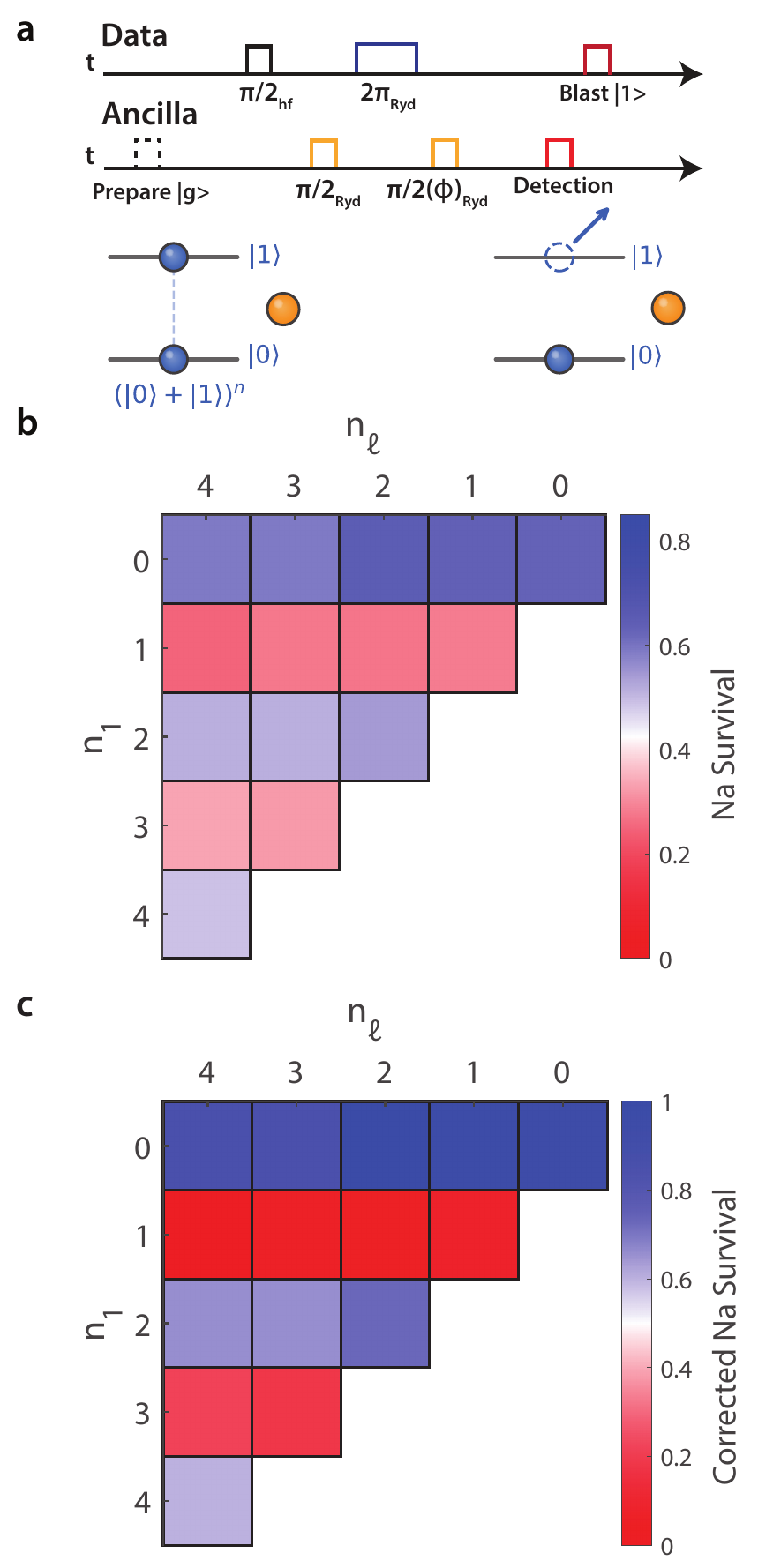}
    \caption{\textbf{Ancilla-based $Z$-parity readout with superposition Cs input states.}
    \textbf{(a)} Pulse sequence for the stabilizer measurement protocol. A global $\pi/2$ hyperfine 
    pulse prepares each loaded Cs atom in $\ket{+} = (\ket{0}+\ket{1})/\sqrt{2}$. A final 
    Cs blast pulse removes atoms in $\ket{1}$, so atoms remaining in $\ket{0}$ are detected.
    \textbf{(b)} Raw data of Na survival probability as a function of $n_{1}$, the number 
    of Cs atoms that projected onto $\ket{1}$, grouped by the number of initially loaded Cs 
    atoms $n_\ell = 0$--$4$. 
    \textbf{(c)} Corrected data by dividing out single-Na Rydberg Ramsey contrast.}
    \label{fig:plusstate}
\end{figure}
We prepare each loaded Cs atom in the superposition state $\ket{+} = (\ket{0}+\ket{1})/\sqrt{2}$ with a global hyperfine $\pi/2$ pulse (Fig.~\ref{fig:plusstate}\textbf{a}). 
Measurement of the stabilizer $Z_1 Z_2 Z_3 Z_4$ should project the many-body state of the Cs atoms into either the odd or even sector. 
To verify this, we carry out the four-atom stabilizer readout scheme as above, but additionally implement state-selective Cs detection by blasting out Cs atoms in $|1\rangle$ before imaging the arrays. 
For $n_\ell$ loaded Cs atoms, we directly measure the number $n_0$ projected into the $|0\rangle$ state via atom survival, and infer $n_1 = n_\ell - n_0$. We then expect the survival or loss of the ancilla to read
\begin{align}
    \Pi_i Z_i = (-1)^{n_1}.
\end{align}
We plot the measured Na survival as a function of $n_\ell$ and $n_1$ (Fig.~\ref{fig:plusstate}\textbf{b}), observing a clear even-odd separation in the Na readout as a function of $n_{1}$. This provides a proof-of-principle demonstration that the fixed-phase ancilla measurement faithfully resolves the plaquette $Z$-parity for arbitrary Cs configurations. 


The fidelity of this stabilizer measurement is limited by three contributing factors. First, as shown in Fig.~\ref{fig_plaq}\textbf{b} and Fig.~\ref{fig_plaq}\textbf{c}, the fringes within a given parity sector exhibit a small phase spread, which we attribute primarily to spatial inhomogeneity across the plaquette. This slightly reduces the compensation of the finite interspecies interaction $V_{\mathrm{NaCs}}$, an effect that would be further suppressed with improved beam homogeneity and tweezer uniformity.


Second, this spatial inhomogeneity, together with the finite fidelity of the Cs $2\pi$ gate ($95\%$), induces an additional per-qubit reduction in the Ramsey contrast. We measure an 18\% reduction in contrast for each additional data qubit in $\ket{1}$ (Fig.~\ref{fig_plaq}\textbf{c,d}). This feature is also visible in the bottom region of Fig.~\ref{fig:plusstate}\textbf{b,c}. A detailed breakdown of these contributions is provided in the Supplementary Material.

Third, the single-Na Rydberg Ramsey decoherence during the Cs $2\pi$ pulse sets the overall contrast envelope shared by all Cs configurations. 
The dominant decoherence mechanisms are likely to include Rydberg laser phase noise, Doppler shifts due to atom motion at finite temperature, and shot-to-shot intensity fluctuations \cite{PhysRevA.97.053803,PhysRevLett.121.123603,Levine2019}, pointing to clear technical paths for improvement.
With a heavier ancilla atom such as Rb and a higher Rabi frequency enabled by more laser power, some of these decoherence effects could be reduced. One could further leverage phase noise-reduction techniques such as active feedforward \cite{hua2025feedforward,denecker2025measurement,chao2024PDH,chao2025robust,li2022active}, as well as implement Raman sideband cooling~\cite{PhysRevA.97.063423,Spence_2022,Kaufman2012,Thompson2013}. This suggests that the single-atom decoherence could be largely mitigated via experimental upgrades, and we would then anticipate a greatly improved contrast as shown in Fig.~\ref{fig:plusstate}\textbf{c}.

\section*{Discussion and Outlook}
We have demonstrated a proof-of-principle Pauli-$Z$-type stabilizer readout on four-atom plaquettes in a dual-species Rydberg array, using a single global Rydberg pulse sequence without atom rearrangement. With the addition of a hyperfine $\pi/2$ pulse on all data atoms both before and after the Rydberg pulse block, the same protocol measures Pauli-$X$-type stabilizers.
In addition to technical improvements, further theoretical work on the design of many-qubit gates could also address and mitigate main error sources and improve this stabilizer measurement fidelity. Whereas two-qubit gate fidelities have improved substantially due to theoretical work improving robustness against common sources of noise, comparable analysis and optimization largely remains unexplored in the multi-qubit case \cite{locher2026multiqubitrydberggatesquantum}.

While this work demonstrates one round of Pauli-$Z$ stabilizer readout, a full implementation of QEC would require multiple rounds of stabilizer readout in both the $X$ and $Z$ bases. A natural next step involves pushing towards a more full-fledged QEC implementation by integrating complementary capabilities such as dual-species atom rearrangement~\cite{PhysRevLett.128.083202}, local light shifts of the ground-Rydberg transition~\cite{bluvstein2022quantum,2022Natur.604..457G,PhysRevA.90.023415,PhysRevX.13.041035}, and fast imaging of the ancilla~\cite{PhysRevA.97.063613,Su:2024urm,n3bg-7yw7}.
The same techniques apply to the measurement of geometrically local stabilizers or gauge operators of variable weight in other QEC code families, allowing natural access to surface codes (weights two and four), color codes (weights three and six), and subsystem codes~\cite{Bombin_2015,PhysRevX.5.031043}. The operation of Clifford-deformed codes~\cite{dua2024clifford} offering tunable resilience to biased noise is also in principle possible with the addition of locally addressable hyperfine $\pi/2$ pulses.

Together with the species-selective control demonstrated here, these ingredients consolidate the four spatial zones typically required for syndrome extraction into a single science zone interfaced with a reservoir zone. This zone consolidation maximizes the fraction of the available field of view devoted to active qubits and reduces the transport overhead of each syndrome-extraction cycle, bringing the dual-species architecture into alignment with the requirements of low-overhead fault-tolerant quantum computation.

\section{Acknowledgements}
We gratefully acknowledge the insights of and fruitful discussions with Harry Zhou, Lysander Christakis, and Bingtian Ye. We thank Yichao Yu and Till Rosenband for contributions to the development of the experimental control system. This work was supported by IARPA and the Army Research Office under the Entangled Logical Qubits program (W911NF-23-2-0219); U.S. Department of Energy, Office of Science, National Quantum Information Science Research Centers, Quantum Systems Accelerator (DE-SCL0000121), and the National Science Foundation via the National Quantum Virtual Laboratory (2533041) and the STAQ (Software-Tailored Architectures for Quantum codesign) program (PHY-2325080). 
N. Y. Y. acknowledges support from a Simons Investigator award.
%

\bibliography{master_ref_Dec2023}

\clearpage

\section{Supplementary Material}
\setcounter{figure}{0}
\renewcommand{\thefigure}{S\arabic{figure}}
\setcounter{equation}{0}
\renewcommand{\theequation}{S\arabic{equation}}
\begin{figure*}[!ht]
   \centering
    \includegraphics[width =.8\linewidth]{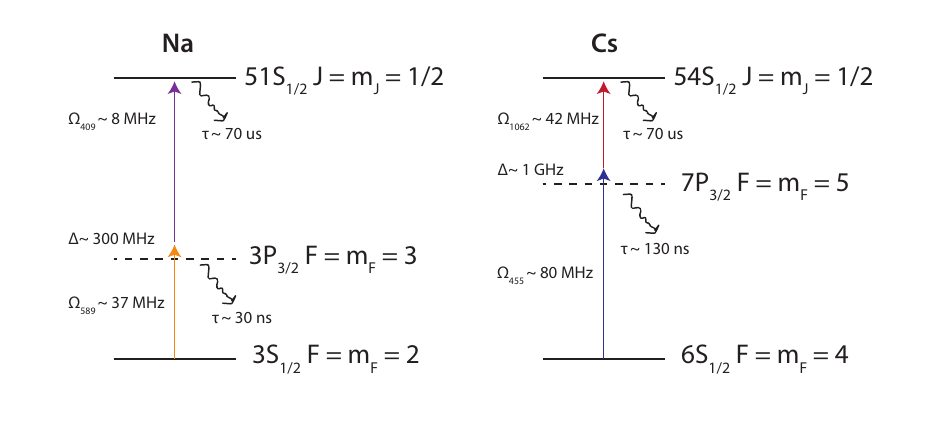}
    \caption{\textbf{Rydberg excitation parameters} We show typical values of the single-photon Rabi frequencies for each species, as well as the Rydberg lifetimes and intermediate state lifetimes.}
    \label{fig:Rydberg_param}
\end{figure*}
\subsection{Experimental system}
\subsubsection{Optical tweezer arrays}
We load arrays of ${}^{23}\mathrm{Na}$ and ${}^{133}\mathrm{Cs}$ atoms into optical tweezer arrays from overlapped 3D magneto-optical traps (MOTs). This choice of atomic species is determined by earlier work on NaCs molecule assembly on the apparatus~\cite{Liu2018,Liu2019,PhysRevX.11.031061}.  We trap Cs at 1064 nm (Precilasers, YFA-SF-1050-50-CW, seed Mephisto) and Na at 616 (Precilaser SFG), which is insensitive to light shifts on the D1 line \cite{Aliyu2021}.  We stochastically load Cs with probability about 50\% into traps of 300 $\mu$K depth. For Na, we apply D1 grey molasses after the D2 MOT stage and load into 1 mK traps with 60-70\% probability.

Each 2D tweezer array is generated by a Hamamatsu spatial light modulator (SLM), with each species having its own SLM suitable for its tweezer wavelength. We achieve trap intensity uniformity $\leq 2\%$ after 4-5 rounds of feedback with the phase-fixed weighted Gerchberg-Saxton (WGS) algorithm. The SLM tweezer for each species is combined on a dichroic and focused through a microscope objective (NA = 0.55).

\subsubsection{Rydberg laser system}
We employ a two-photon excitation scheme with counterpropagating beams for coherent Rydberg excitation on each species as illustrated in (SM Fig~\ref{fig:Rydberg_param}). For Cs, the 455 nm light is supplied by a frequency-quadrupled fiber laser system (Precilaser, FL-SF-455-1-CW) outputting 800 mW at 455 nm. This is passed through an acousto-optical modulator (AOM) in a single-pass configuration (AA Optoelectronics, MQ180-A0, 25-VIS) and fiber-coupled to the experiment. The 1062 nm light is supplied by an external cavity diode laser (ECDL, Timebase ECQDL-200FC). This laser is stabilized to an ultra-low-expansion (ULE) cavity (see below), and the cavity-filtered transmission is used to injection lock another diode (EYP-RWL-1060-00100-1300-SOT01-0000). This diode seeds a fiber amplifier (Precilaser YFA-SF-1050-50-CW) which outputs up to 50 W at 1062 nm. The output of the fiber amplifier is passed through a single-pass AOM and fiber-coupled to the experiment. Both lasers are intensity stabilized with home-built analog PID servos on each cycle of the experiment. The 455 nm light is additionally equipped with an analog sample-and-hold to allow for intensity stabilization on a shot-to-shot basis.

For Na, the 589 nm light is generated by an 1178 nm ECDL (Toptica DL Pro) which seeds a semiconductor optical amplifier (SOA) (Innolume SOA1190090YY20DBXXXX), generating up to 200 mW of light at 1178 nm. The SOA pumps a fiber-coupled waveguide second harmonic generation (SHG) module (NTT WH-0589-000-A-B-C) with a linear efficiency of $35-40\%$ for a total output of 70-80 mW at 589 nm. This 589 nm light is passed through a double-pass AOM (Gooch and Housego, R23080-1-LTD) and fiber coupled to an intermediate combination board, where it is merged with optical pumping (OP) and state detection light (also at 589 nm) before being fiber-coupled onto the experiment. It is intensity stabilized with a home-built Arduino PID servo with digital sample-and-hold. For delivery onto the atoms, the 589 nm light is focused to a beam waist of 300 $\mu$m. The 409 nm light fundamental is generated by a Ti:Sapphire laser (M Squared Solstis) at 818 nm, which is then cavity frequency-doubled to 409 nm (M Squared ECD-X). The light is coupled into a photonic crystal fiber (PCF, Alphanov LMA-PM-5) and focused to a beam waist of 30-50 $\mu$m with an adjustable-focus collimator (Thorlabs C40APC-A). 

All four Rydberg lasers are stabilized to a common ULE cavity (Notched cavity from Stable Laser Systems). For visible-wavelength lasers, the IR fundamental is locked to the cavity. The cavity finesse is measured to be $\mathcal{F} = 24000$ at 911 nm and $\mathcal{F} = 26000$ at 1062 nm. 

The Rydberg beam parameters are also shown in (SM Fig.~\ref{fig:Rydberg_param}). We note that the imbalance of the single-photon Rabi frequencies on Na is due to substantial reflection of 409 nm light on each surface of our double-side coated glass cell. The potential etalon effect arising from this reflection is mitigated by aligning the 409 nm beam at a modest incidence angle; we choose  $2^\circ$ due to geometrical constraints. Nonetheless only $32\%$ of the total 409 nm power is transmitted through the glass. In a future upgrade to our apparatus, we plan to use an uncoated glass cell to overcome this limitation.

\subsubsection{State preparation, state-selective detection, and imaging}
\begin{figure}[!ht]
   \centering
    \includegraphics[width =1\linewidth]{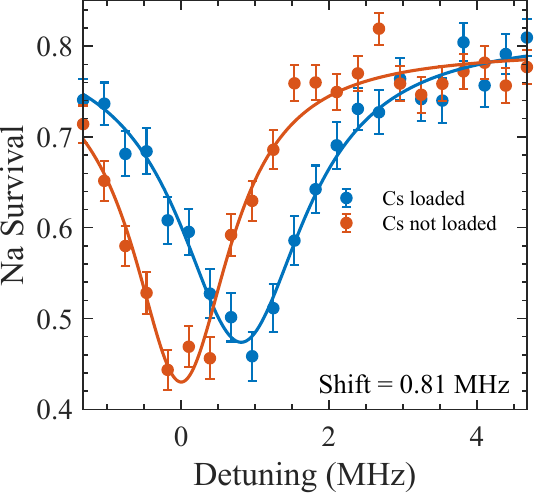}
    \caption{\textbf{Interspecies Rydberg-Rydberg interaction.}  A typical interspecies interaction measured by Rydberg resonance shift of Na conditioned on whether a nearby Cs Rydberg atom is loaded.}  
    \label{fig:interaction}
\end{figure}
We optically pump Cs to $|F,m_F\rangle = |4,4\rangle$ and Na to $|F,m_F\rangle = |2,2\rangle$ using $\sigma^+$-polarized light. For Cs, we address the $|3,3\rangle\to|4',4'\rangle$ and $|4,4\rangle\to|4',4'\rangle$ transitions on the $D_2$ line, since the $|5',5'\rangle$ cycling transition is detuned by 251 MHz and does not adversely affect the OP fidelity. For Na, we address the $|1,1\rangle\to |2',2'\rangle$ transition on the $D_2$ line, but the $|2,2\rangle\to|2',2'\rangle$ transition on the $D_1$ line. This means that the $m_F$-level pumping does not suffer from off-resonant scattering through $|3',3'\rangle$ on the $D_2$ line, which is only detuned by 58 MHz. During OP, we modulate the Na trap with a square wave at 2.5 MHz and 80\% on-time duty cycle. We calibrate the OP intensity and detuning to pump atoms during the tweezer off-time, such that the OP fidelity and time do not suffer from light shifts on the non-magic $D_2$ transition.  

We optionally perform hyperfine state-selective detection by inserting a resonant 'blast' pulse on the cycling transition $|4,4\rangle\to|5',5'\rangle$ (for Cs) or $|2,2\rangle\to|3',3'\rangle$ (for Na) which converts atoms in the $|4,4\rangle$ or $|2,2\rangle$ states into loss. This enables us to calibrate the OP fidelity by driving a resonant microwave or Raman pulse from $|4,4\rangle\to|3,3\rangle$ or $|2,2\rangle\to|1,1\rangle$ followed by state-selective blast. Since population which is successfully transferred to the lower hyperfine state survives the blast, the fit contrast of this pulse gives a lower bound on state preparation fidelity (accounting separately for decoherence by including an exponential decay). Using this method, we find over OP fidelity of $95\%$ for both species.

We image Cs with red-detuned light on the $D_2$ cycling transition through the 3D MOT beams with a survival probability of 97.4\% between two successive images. For Na, imaging on the $D_2$ line requires the tweezer and imaging light to be strobed out of phase to avoid repulsive light shifts from the excited states~\cite{Hutzler2017}. Since this reduces the time-averaged tweezer power and thereby limits the number of tweezers that can be generated at sufficiently high trap depth, we instead opt for imaging on the $D_1$ line, and the tweezer wavelength is chosen to cancel the differential light shift between the ground and excited states during cooling and imaging. We use a pair of counterpropagating beams focused to $\sim100$ $\mu$m waist to deliver blue-detuned light on the $D_1$ line, tuned to optimize for 1D grey molasses cooling along a radial direction of the tweezer~\cite{Aliyu2021,Colzi2016,PhysRevX.9.011057}. Survival between successive images is limited to 89.1\%, likely limited by atom heating during the image. Further work is required to identify the sources of, and mitigate, this loss.

\subsection{Dual-species interaction spectroscopy}
In this experiment, we choose Rydberg states 
\begin{align*}
    |r\rangle_\text{Cs} &= |54S_{1/2}\rangle\\
    |r\rangle_\text{Na} &= |52S_{1/2}\rangle.
\end{align*}
The Cs Rydberg state, which we keep unchanged from our prior work~\cite{doi:10.1126/science.adq0278}, was chosen to achieve reasonably high Cs-Cs Rydberg interactions ($C_\text{6,intra} = 27.96~\text{GHz}~\mu\text{m}^6$) without suffering from detuning drift due to electric field noise. The Na Rydberg state is chosen to achieve maximal interspecies interaction strength with the fixed Cs state; we calculate $C_{6,\text{inter}} = 65.7~\text{GHz}~\mu\text{m}^6$ using the Alkali Rydberg Calculator (ARC) Python package~\cite{Sibalic2017}. 

To measure the interspecies interaction strength, we arrange Na and Cs atoms into pairs. We first excite one atom of the pair to its Rydberg state and subsequently probe the Rydberg resonance of the other by scanning the excitation laser frequency. A typical interaction measurement for a pair of Na-Cs in shown in  Fig.~\ref{fig:interaction}.

\subsection{Two-atom stabilizer readout protocol}
Here we detail the operation of the stabilizer readout protocol in the case of two atoms, as discussed in the main text. 
\begin{enumerate}
    \item The data atom begins in $|\psi\rangle_\text{data} = a|0\rangle_\text{data} + b|1\rangle_\text{data}$. The ancilla is prepared in $|g\rangle_\text{ancilla}$. 

    \item A ground-Rydberg $\pi/2$ pulse is applied to the ancilla, mapping it to $\frac{1}{\sqrt{2}}\left(|g\rangle_\text{ancilla}+i|r\rangle_\text{ancilla}\right)$.

    \item A ground-Rydberg $2\pi$ pulse is applied to the data qubit. In the interspecies blockade regime, the ancilla blocks Rydberg excitation of the data qubit when in the state $|r\rangle_\text{ancilla}$. By contrast, when the ancilla is in the state $|g\rangle_\text{ancilla}$, the data qubit undergoes a full resonant ground-Rydberg $2\pi$ rotation and acquires a global phase of $-1$. The state of the system is then
    \begin{align*}
        &\frac{1}{\sqrt{2}}\big[|g\rangle_\text{ancilla}\otimes\left(a|0\rangle_\text{data}-b|1\rangle_\text{data}\right)\\&+i|r\rangle_\text{ancilla}\otimes\left(a|0\rangle_\text{data}+b|1\rangle_\text{data}\right)  \big]
    \end{align*}
    which can be rewritten as 
    \begin{align*}
        &\frac{1}{\sqrt{2}}\big[a|0\rangle_\text{data}\otimes\big(|g\rangle_\text{ancilla}+i|r\rangle_\text{ancilla} \big)\\
        &+b|1\rangle_\text{data}\otimes\big(-|g\rangle_\text{ancilla}+i|r\rangle_\text{ancilla} \big)\big].
    \end{align*}
    In other words, the ancilla superposition acquires a relative $\pi$ phase conditional on the state of the data qubit.

    \item A final ground-Rydberg $\pi/2$ pulse is applied to the ancilla. The state of the system is
    \begin{align*}
        \left(a|0\rangle_\text{data}|r\rangle_\text{ancilla} + b|1\rangle_\text{data}|g\rangle_\text{ancilla} \right).
    \end{align*}

    \item The ancilla is measured in the ground-Rydberg basis. The $|r\rangle_\text{ancilla}$ state is detected as ancilla atom loss due to tweezer anti-trapping, while the $|g\rangle_\text{ancilla}$ state is detected as ancilla atom survival. This projectively measures the data atom into $|0\rangle_\text{data}$ with probability $|a|^2$ or into $|1\rangle_\text{data}$ with probability $|b|^2$. 
\end{enumerate}

\subsection{Interaction-compensated CZ protocol}
\subsubsection{Effects of imperfect blockade on a resonantly-driven $CZ$ gate}
\begin{figure*}[!ht]
   \centering
    \includegraphics[width =.8\linewidth]{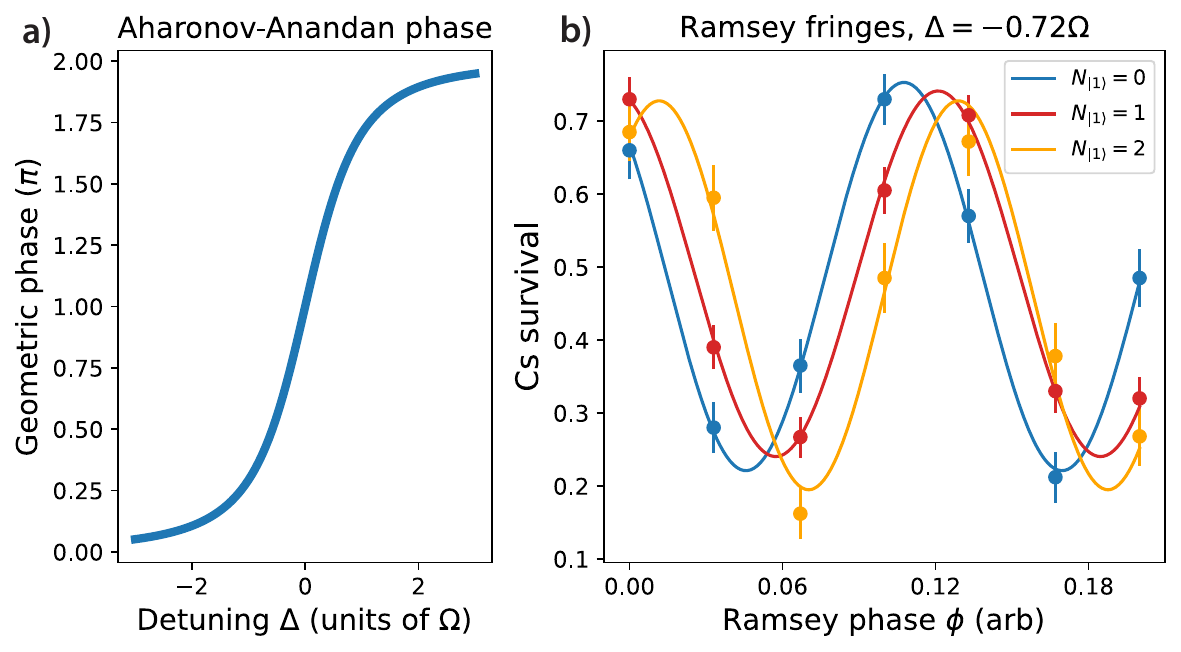}
    \caption{\textbf{Aharonov-Anandan phase.} \textbf{(a)} The Aharonov-Anandan phase plotted for detuning $\Delta = -2~\Omega$ to $\Delta = 2~\Omega$. On resonance, the acquired phase shift is $\pi$. \textbf{(b)} Data taken at $\Delta = -0.72~\Omega$. The acquired phase shift for each additional data atom is between $0.25\pi$ and $0.5\pi$.}
    \label{fig:Aharonov}
\end{figure*}
On dual-species atom plaquettes, there is an important tradeoff between data-ancilla distance and data-data distance. While closely spaced plaquettes can increase the data-ancilla interaction $V_\text{da}$ and situate the experiment further into the blockade regime, this also increases the data-data interaction $V_\text{dd}$, leading to undesirable $n_\text{data}$-dependent shifts of the data ground-Rydberg resonance. In a square geometry, $V_\text{da}$ is automatically enhanced relative to $V_\text{dd}$ by a geometric factor of 8, with further enhancement possible by judicious choice of Rydberg state on each species. In our experiment, the choice of $n_\text{Cs} = 54$ and $n_\text{Na} = 52$ enhances the interspecies $C_6$ coefficient relative to the Cs intraspecies $C_6$ coefficient by a factor of 2.5 and the geometrical factor further increase the $V_{da}/V_{dd}$ up about 20. While further enhancement is in principle possible by tuning to Förster resonance with electric field or choosing Rydberg states with even stronger $C_6$ coupling (normally with both species at higher principal quantum number), it remains difficult to simultaneously satisfy (a) the interspecies blockade constraint to ensure proper operation of the blockaded $CZ$ gate and (b) the constraint of low $V_\text{dd}$ to avoid data atom crosstalk. 

We first briefly discuss the nature of the errors accrued when attempting blockaded $CZ$ gates outside of the blockade regime, beginning with a two-atom system in the state
\begin{align*}
    |\psi\rangle_\text{ancilla}|\psi\rangle_\text{data} = \frac{1}{\sqrt{2}}\left(|g\rangle+|r\rangle\right)_\text{ancilla}\otimes|1\rangle_\text{data}
\end{align*}
The data atom is driven resonantly ($\Delta = 0$) with Rabi frequency $\Omega$ and interacts with the ancilla with strength $0<V_\text{da}<\infty$. With the ancilla in $|g\rangle$, the data atom undergoes an ideal resonant $2\pi$ pulse and accrues a phase of -1 on this part of the wavefunction. The issue arises when the ancilla is in $|r\rangle$. When $\Omega/V\sim 1$, the dynamics of the data atom are altered but not blockaded: it undergoes detuned dynamics under the enhanced Rabi frequency $\Omega_\text{eff} = \sqrt{\Omega^2 + V_\text{da}^2}$ for time $t_{2\pi}= 1/\Omega$, leaving the two-atom system in the state
\begin{align*}
    &\frac{1}{\sqrt{2}}\big(-|g\rangle_\text{ancilla}\otimes|1\rangle_\text{data} \\&+ |r\rangle_\text{ancilla}\otimes \exp\left[-\frac{iH(\Omega,\Delta)t}{\hbar} \right]|1\rangle_\text{data}\big)
\end{align*}
where
\begin{align*}
    H(\Omega,V_\text{dd}) = \left(\begin{array}{cc}0&\Omega/2\\\Omega/2&V_\text{da} \end{array}\right).
\end{align*}
The data atom does not undergo a closed trajectory on its Bloch sphere: it is left with non-negligible population in the Rydberg state so that it cannot be said to have accumulated a geometric phase, and suffers population loss from anti-trapping.

In the scenario of $\Omega/V_\text{da}\ll 1$, the population dynamics of the data atom are suppressed by the ancilla blockade, but phase error is still accrued whenever $V_\text{da}<\infty$. To first order the data atom acquires an AC Stark shift from the excitation beam and accrues a dynamical phase $\varphi_\text{dyn}$
\begin{align*}
    |1\rangle\to e^{-i|\Omega|/V_\text{da}}|1\rangle
\end{align*}
This results in a relative phase on the ancilla which deviates from $\pi$:
\begin{align*}
    \delta\varphi&=\varphi_\text{A-A}(0,\Omega) - \varphi_\text{dyn}(-V,\Omega,t=1/\Omega)\\
    &\sim\pi-\frac{|\Omega|}{V_\text{da}}.
\end{align*}
This phase error can be made arbitrarily small in the perfectly blockaded limit $\Omega/V_\text{da}\to 0$, but in practice is difficult to achieve ratios lower than $\Omega/V_\text{da} = 1/50$  in dual-species arrays of alkali atoms without also increasing $V_\text{dd}$ towards impractically large values. In other words, the performance of a $CZ$ gate on dual-species plaquettes is limited by the requirement to maintain the hierarchy of scales $V_\text{dd}\ll\Omega\ll V_\text{da}$.

\subsubsection{Aharonov-Anandan phase}

To further characterize phase errors accrued in the $CZ$ gate protocol, it is important to understand the detuning-dependent global phases which can result from an atom's interaction with detuned light. In the case discussed above with $\Omega/V_\text{da}\ll 1$, the accumulated phase due to the far-detuned AC Stark effect is a dynamical phase depending on the total evolution time. When the driving field is close enough detuned that there are nontrivial population dynamics, the atom can accumulate a geometric phase when it undergoes a closed trajectory on its Bloch sphere. This phase is known as the Aharonov-Anandan phase, and it is a geometric phase in the sense that its value depends on the trajectory of the Bloch vector. A well-known example is the global -1 phase acquired by any spin-1/2 system when driven with a resonant $2\pi$ pulse. 

The Aharonov-Anandan phase for Rabi frequency $\Omega$ and detuning $\Delta$ is between $0$ and $2\pi$ and depends on the surface area swept out on the Bloch sphere:
\begin{align}
    \label{AAphase}
    \varphi_\text{A-A}(\Delta,\Omega) = -\pi\left(1+\frac{\Delta}{\sqrt{\Omega^2 + \Delta^2}} \right).
\end{align}
We plot this phase for $\Omega =1$ in (Fig.~\ref{fig:Aharonov}a). In (Fig.~\ref{fig:Aharonov}b) we demonstrate the effect of the Aharonov-Anandan phase for $\Delta < 0$ on the measured ancilla Ramsey fringe.The data atom $2\pi$ time is adapted to account for the detuned effective Rabi frequency, ensuring a closed trajectory. When $\Delta < 0$, the ancilla acquires a phase $\delta\varphi_\text{ancilla}<\pi$. This highlights the importance of resonance stability when driving resonant gates. Phase errors can accumulate due to e.g. intensity fluctuations of either Rydberg beam in the two-photon excitation, which give rise to fluctuating AC Stark shifts.

\subsubsection{Interaction-compensated protocol}
\begin{figure*}[!ht]
   \centering
    \includegraphics[width =.8\linewidth]{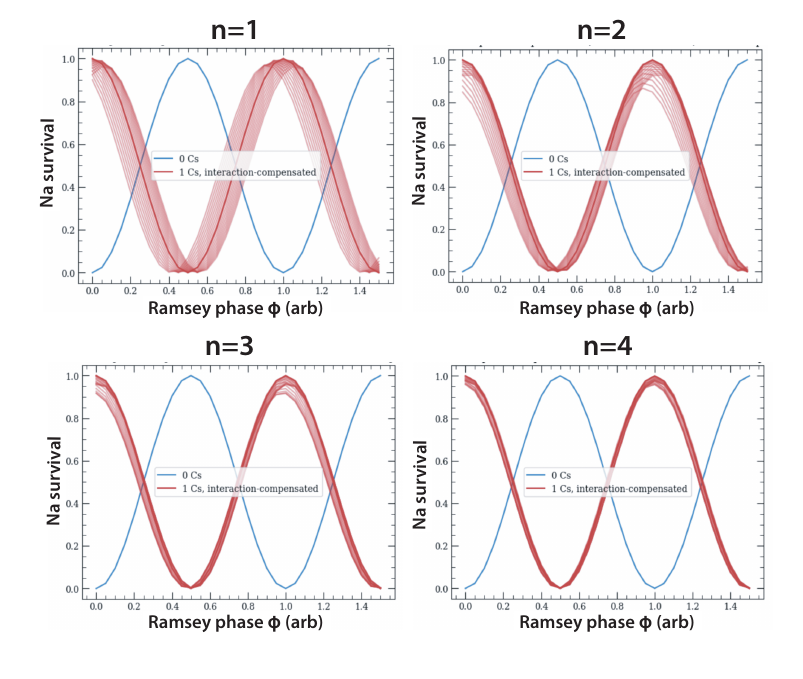}
    \caption{\textbf{Simulated interaction compensated protocol} Dependence of the measured ancilla phase under the interaction-compensated protocol. Each subplot shows the ancilla Ramsey fringe shift in a two-atom experiment for fixed $V$ and varying $n$. The ideal case where $V$ is a single fixed value always shows a perfect $\pi$ phase shift (dark red). Variations of the fringe for $\pm 20\%$ fluctuations of $V$ are shown in lighter red. For higher $n$, the values of $\Omega$ and $\Delta$ approach those of the resonant blockaded gate, and the protocol is more resilient against fluctuation of $V$.}
    \label{fig:Simulated interaction compensated protocol}
\end{figure*}
This ability to tune the Aharonov-Anandan phase also provides an additional degree of freedom that can be used to directly compensate the effects of finite $V_\text{da}$. This approach was proposed in \cite{Shi2017} and further explored in \cite{cole2026asymmetricfastrydberggate}. The idea of using the detuning to compensate for finite interaction errors has also been discussed in the context of other gates, e.g. the Levine-Pichler $CZ$ gate \cite{Levine2019}. Here we expand on the discussion in the main text. We drive the atom with Rabi frequency $\Omega$ and detuning $\Delta$. When the ancilla is in $|g\rangle$ there is no further effect from $V_\text{da}$, so the data atom undergoes dynamics at detuning $\Delta$ and effective Rabi frequency $\Omega_\text{non-interacting} = \sqrt{\Omega^2 + \Delta^2}$. When the ancilla is in $|r\rangle$, the data atom ground-Rydberg resonance is shifted upward by $V_\text{da}$, such that it undergoes dynamics at detuning $\Delta-V_\text{da}$ and effective Rabi frequency $\Omega_\text{interacting} = \sqrt{\Omega^2 + (\Delta-V_\text{da})^2}$. To ensure that the data atom undergoes a closed trajectory on the Bloch sphere in each case, we first require the condition
\begin{align}
    \label{phase1}
    \sqrt{\Omega^2 + (\Delta-V_\text{da})^2} = n\sqrt{\Omega^2 + \Delta^2}, n\in\mathbb{Z}.
\end{align}
Next, we require that the differential phase between the interacting and non-interacting cases is $\pi$:
\begin{align}
    \label{phasediff}
    \varphi_{A-A}(\Delta,\Omega) - \varphi_\text{A,A}(\Delta-V_\text{da},\Omega) = \pi.
\end{align}
Substituting the explicit expression (Eq. \ref{phasediff}) into (Eq. \ref{AAphase}) gives
\begin{align*}
    &-\pi\left(1+\frac{\Delta}{\sqrt{\Omega^2+\Delta^2}}\right) \\&+n\pi\left(1+\frac{\Delta-V_\text{da}}{\sqrt{\Omega^2+(\Delta-V_\text{da})^2}}\right) = \pi.
\end{align*}
The additional factor of $n$ in the second term reflects the fact that when $\Omega_\text{interacting} = n\Omega_\text{non-interacting}$, it additively acquires the same geometric phase $n$ times. With the use of (Eq. \ref{phase1}) we then find
\begin{align}
    n &= \frac{\sqrt{\Omega^2+\Delta^2}}{2\Delta}\\
    V &= \frac{\Omega^2 + \Delta^2}{2\Delta} = n\sqrt{\Omega^2+\Delta^2}.
\end{align}
This can be rearranged to give solutions $\Omega, \Delta$ given $V$ and $n$:
\begin{align}
    \Delta &= \frac{V}{2n^2}\\
    \Omega &= \frac{V}{2n^2}\sqrt{4n^2-1}.
\end{align}
The solution for $n = 1$ is $\Delta = V_\text{da}/2$ and $\Omega = (\sqrt{3}/2)V_\text{da}$. Since $\Omega/ V_\text{da}\sim0.87 $, we can now operate the gate under the less-restrictive condition $V_\text{dd}\ll\Omega\sim V_\text{da}$, both acquiring the correct ancilla phase and also allowing for weaker data-data blockade. 

The implementation of this protocol of course depends on measuring $V$, so a natural question arises as to how robust it is to fluctuations of $V$, both static and dynamic. To shed light on this, (SM Fig.~\ref{fig:Simulated interaction compensated protocol}) shows two-atom simulations of ancilla Ramsey fringes for different values of $n$, setting $\Omega = 1$. We see that as $n$ is increased, the protocol becomes increasingly robust against fluctuations of $V$. This has the intuitive explanation that as $n\to\infty$, we also have $\Omega,\Delta\to 0$, meaning that the gate asymptotically becomes a resonant gate with the strong blockade condition $\Omega/V\to 0$. However, larger values of $n$ increase the total gate time, leading to further dephasing of the ancilla (see below), as well as reducing $\Omega$ relative to $V_\text{dd}$, making the gate more susceptible to the effects of data-data interactions. In practice these tradeoffs can be empirically optimized and are more than likely subject to other experimental constraints, for instance in the achievable maximum $\Omega$. Our guiding heuristic is to minimize $n$ in order to maximize $\Omega$ within experimental limits.

\subsection{Stabilizer measurement error analysis}
\begin{figure*}[!ht]
   \centering
    \includegraphics[width =.8\linewidth]{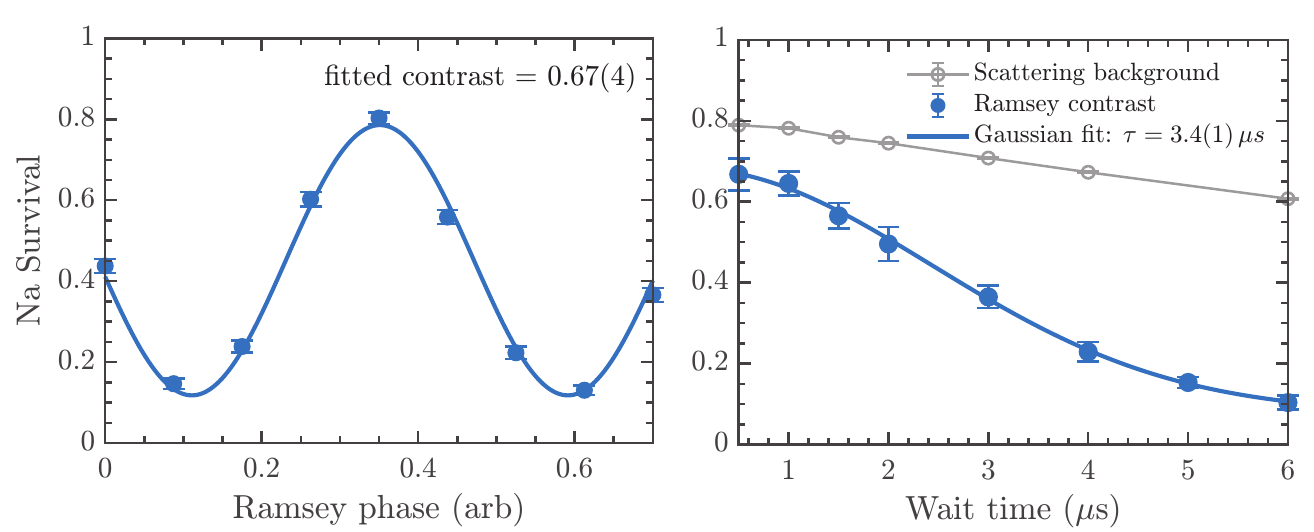}
    \caption{\textbf{Na ancilla Ramsey coherence.} (Left) An example phase Ramsey fringe measured at a wait time of $1~\mu$s. (Right) Fitted Ramsey fringe contrast as a function of wait time. The pulse sequence follows Fig.~\ref{fig:twoatom}\textbf{e}, with the delay between the two $\pi/2$ pulses varied. The scattering background (open markers) is obtained from the same sequence with the 409~nm Rydberg excitation laser turned off.}
    \label{fig:Ramsey}
\end{figure*}
Both the amplitude and phase of the ancilla Ramsey fringe can be negatively affected by a variety of single-atom and multi-atom error sources. Here we list and characterize the most important sources of error for the readout scheme.  
\subsubsection{Single-atom sources of error}
\begin{enumerate}
    \item \textbf{Ancilla ground-Rydberg $T_2^*$ dephasing}. During the data atom $2\pi$ pulse, the ancilla atom is idled in the ground-Rydberg superposition state
    \begin{align*}
        |\psi\rangle_\text{ancilla} = \frac{1}{\sqrt{2}}\left(|g\rangle+i|r\rangle \right).
    \end{align*}
    Figure (SM Fig.~\ref{fig:Ramsey}\textbf{a}) shows an example single-atom Na Ramsey fringe. The phase coherence of this superposition is limited by a variety of homogeneous and inhomogeneous broadening sources, including Rydberg laser phase noise, intermediate-state scattering during Rydberg excitation, and Doppler shifts due to finite atom temperature. We measure a $T_{2}^{*}$ coherence time of 3.4 $\mu$s on Na characterizing the Gaussian decay of the ground-Rydberg Ramsey fringe amplitude as $\exp(-[t/T_{2}^{*}]^2)$. If the scattering background loss is divided out, the decay time is 3.7 $\mu$s. This sets the amplitude of the $n_\text{data} = 0$ Ramsey fringe and therefore the maximum achievable detection contrast. We find that after a point, this decay timescale is rather insensitive to Rydberg laser PID lock parameters and intermediate-state scattering rate, likely indicating that atom temperature is the primary source of decoherence. We plan to address this by implementing 3D Raman sideband cooling of Na, rendered more difficult by the high Lamb-Dicke parameter of Na ($\eta_\text{axial}\approx0.68$).
    
    Because the total ancilla idle time is set by the data atom Rabi frequency $\Omega$, it is not always advantageous to improve the interspecies blockade by reducing $\Omega$.
    
\item \textbf{Ancilla Rabi coherence time $\tau$}. We give exemplary Na ground-Rydberg Rabi oscillations in main text (Fig.~\ref{fig:Exp_scheme}\textbf{d}), showing exponential dampening at a rate of about $\tau = 12$ $\mu$s. This is limited primarily by intermediate-state scattering and secondarily by laser phase noise, likely on the $1178~nm$ ECDL. The primary effect of Rabi decoherence is to limit the contrast of the Ramsey fringe, setting another limit on the maximum achievable detection fidelity. This can be further improved by (a) increased power on the upper leg of the two-photon excitation at $409 ~\text{nm}$, and (b) phase noise feedforward on the lower leg at $589~\text{nm}$. 
    
    \item \textbf{Ancilla scattering loss}. For Na, an additional complication is ground-state atom loss due to intermediate-state scattering during Rydberg excitation, made worse (relative to heavier alkali atoms) by the higher recoil velocity of Na. We operate in an intensity regime where this additional source of scattering loss is not worse than $2-3\%$ on top of the baseline imaging loss. This could be improved with increased 409 nm laser power, allowing operation at similar two-photon Rabi frequencies with reduced intermediate-state scattering on the lower leg.

    \item \textbf{Data Rabi coherence time $\tau$}. We observe exponential decay of Cs ground-Rydberg Rabi oscillations at a timescale of $\tau = 22$ $\mu$s, limited by intermediate state scattering and Rydberg lifetime. This limits the data $2\pi$ pulse fidelity to $95.1\% $. Failure to complete a full $2\pi$ rotation leaves residual data atom population in the Rydberg state, leading to tweezer anti-trapping and atom loss. We partially mitigate this error source by postselecting on the survival of the Cs atom when state-selective pushout is not required. Future work to increase the 1062 nm single-photon Rabi frequency can improve this.
\end{enumerate}

\subsubsection{Multi-atom sources of error}
\begin{enumerate}
    \item \textbf{Finite data-ancilla interaction}. We refer to the main text (Section 2) and supplementary material (above) for extensive discussion of this error source. 

    \item \textbf{Nonzero data-data interaction}. On a plaquette with side length $\ell$, the data atoms interact with strength
    \begin{align*}
        V_\text{dd} = \frac{C_{6,\text{data}}}{\ell^6}.
    \end{align*}
    This leads to shifts of the data atom ground-Rydberg which depend on the number of neighbors also excited:
    \begin{align*}
        \delta_\text{data} = n_\text{nearest}\frac{C_{6,\text{data}}}{\ell^6} + n_\text{next-nearest}\frac{C_{6,\text{data}}}{(\sqrt{2}\ell)^6}
    \end{align*}
    where $n_\text{nearest}\in[0,1,2]$ is the number of nearest-neighbor data atoms starting in $|1\rangle$ and $n_\text{next-nearest}\in[0,1]$ is the number of next-nearest-neighbor (i.e. diagonally opposite) data atoms starting in $|1\rangle$. This effect is mitigated by (a) working on plaquettes of larger $\ell$ and compensating the reduced interspecies interaction in the gate protocol, and (b) working at higher data Rabi frequency $\Omega_\text{Cs}$. To ensure that the distance between data atoms is largest, it is important to confirm that the arrays for each species are in the same focal plane. We confirm this by varying the magnitude of the defocus Zernike correction applied to the Na tweezer array and measuring the resulting interspecies interaction strength; focal alignment corresponds to a clear maximum.

\end{enumerate}

\subsection{Dynamics of data atoms loaded in $|0\rangle$}
For data taken on plaquettes of up to 4 data atoms, we use non-loaded atoms as a proxy for atoms loaded in the $|0\rangle$ state. This dramatically improves our data rate by allowing us to include experimental sequences where all four atoms are not loaded in our averaging. An alternative protocol would involve rearrangement into defect-free plaquettes of four atoms followed by local Raman addressing to flip a variable number of data atoms from $|1\rangle$ to $|0\rangle$ before stabilizer readout. While we do have the ability to rearrange, we currently lack local Raman addressing capabilities, rendering this approach unfeasible on the current apparatus.

\begin{figure}[!ht]
   \centering
    \includegraphics[width =1\linewidth]{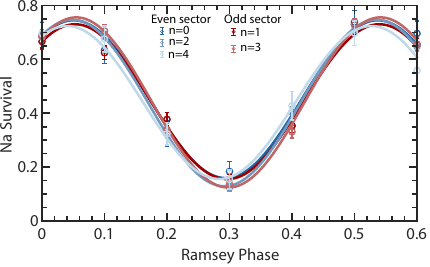}
    \caption{Ramsey fringes measured on the Na ancilla for different numbers of loaded Cs atoms, $n=0$ to $4$, using the interaction-compensation protocol. All the loaded Cs atoms are initialized in $|0\rangle$. All data in both the even-parity (blue) and odd-parity (red) sectors collapse. Error bars denote statistical uncertainties. }
    \label{fig:plaq0}
\end{figure}

We verify that non-loaded atoms behave analogously to atoms loaded in $|0\rangle$ by loading all atoms in the $|+\rangle$ state, running the stabilizer measurement protocol, and including only sequences with data atoms projectively measured into $|0\rangle$. In this case, we expect to see no $n_\text{Cs}$-dependence in the phase Ramsey fringe measured by the ancilla: all fringes should overlap with $n_\text{Cs} = 0$. Figure (SM Fig.~\ref{fig:plaq0}) shows that this is indeed the case.

\subsection{Bootstrap error analysis of data}
Error bars for fits presented in Figs. 2-4 are calculated using bootstrap methods~\cite{bootstrap_Ana}. Briefly, for an experiment with $N$ experimental shots, we sample from the data $N$ times with replacement, calculate the average survival, and perform a fit. This fitting procedure is repeated 300 times. For each fitting parameter, we report the mean over the 300 realizations and extract the bootstrap standard deviation $\xi_\text{b}$. We then fit the mean data, extract the fitting error $\xi_\text{f}$ for each parameter, and report the average error $\xi = \sqrt{\xi_\text{b}^2 + \xi_\text{f}^2}$.



\end{document}